# Vapor pressure dependence of spectral width of EIT in Λ-levels cesium molecular system

H. Chen[1*], H. Li[1], Y. V. Rostovtsev[1], M. A. Gubin[2], V. A. Sautenkov[1,2], and M. O. Scully[1,3]

[1] Department of Physics and Institute for Quantum Studies, Texas A&M University, College Station, TX 77843, USA

[2] P.N. Lebedev Physical Institute, 53 Leninsky pr., Moscow 119991, Russia

[3] Department of Mechanical and Aerospace Engineering, Princeton University, Princeton, NJ 08544, USA

*Corresponding author. e-mail: huichen06@gmail.com

**Abstract**

We have studied electromagnetically induced transparency (EIT) in diatomic cesium molecules in a vapor cell by using tunable diode lasers. We have observed a sub-natural Λ-resonance in absorption molecular band $B^1\Pi_u$ - $X^1\Sigma_g^+$ at different cesium vapor pressures. The width of the EIT resonance shows a linear dependence on cesium vapor pressure. Narrow Λ-resonances in molecules can be used as frequency references for femtosecond laser frequency combs.

**1. Introduction**

Optically induced long lived ground state coherence in three-level atomic and molecular systems can reduce absorption in a very narrow spectral window. This effect is known as electromagnetically induced transparency (EIT) in the Λ-levels quantum system [1]. The corresponding steep dispersion allows a slow propagation of light pulses [2-3] and can be used for modification of the phase-matching conditions for four-wave mixing [4]. Narrow Λ-resonances in the absorption are used in atomic clocks [5] and atomic magnetometers [6]. Recently the EIT resonances have been observed in lithium [7] and potassium [8] molecules in cascade schemes, and in acetylene molecules in Λ-levels scheme [9]. Slow-light effects have been demonstrated in [9]. In current paper we present observation of a narrow Λ -resonance in the diatomic cesium molecules and suggest application of the Λ-resonances in a molecular medium as frequency references for time and frequency standards.

**2. Setup and obtained results**

We have observed a sub-natural Λ-resonance in the absorption cesium molecular band $B^1\Pi_u$ - $X^1\Sigma_g^+$ at 780 nm. The absorption lines in this molecular band cover wavelength region from 755 nm to 810nm [10]. Spectral intervals between the lines are comparable with Doppler broadening [11]. By using high-resolution laser spectroscopy the molecular ground states were mapped in [12 - 15]. These published results helped us to identify the coupled ground states in our experiment. We have used two external cavity diode lasers (ECDL) to produce coherence between rotational-vibrational (rovibronic) ground states. Schematic of the experimental set-up and simplified Λ-scheme of energy levels are presented in Fig. 1. Output power of each laser was several mW. Linear polarized laser beams with the same optical power were combined by beam splitter and sent to glass cell with a drop of metal cesium. The divergence of the beams and the angle between beams were less than $10^{-2}$ rad. The beam diameter is of 0.1 cm, the cell length is of 7.5 cm. The cell is installed in magnetic shield and can be heated till 540 K. The cesium vapor at a high temperature is combination of cesium atoms and cesium diatomic molecules (dimers) and defined by the cell temperature [16]. The frequency dependence of the cell transmission was recorded by photo-detector (PD) and digital oscilloscope connected to a computer. The frequency tuning was controlled by a diffraction spectrometer (resolution 0.1 nm) and by a confocal reference cavity (resolution 10 MHz).

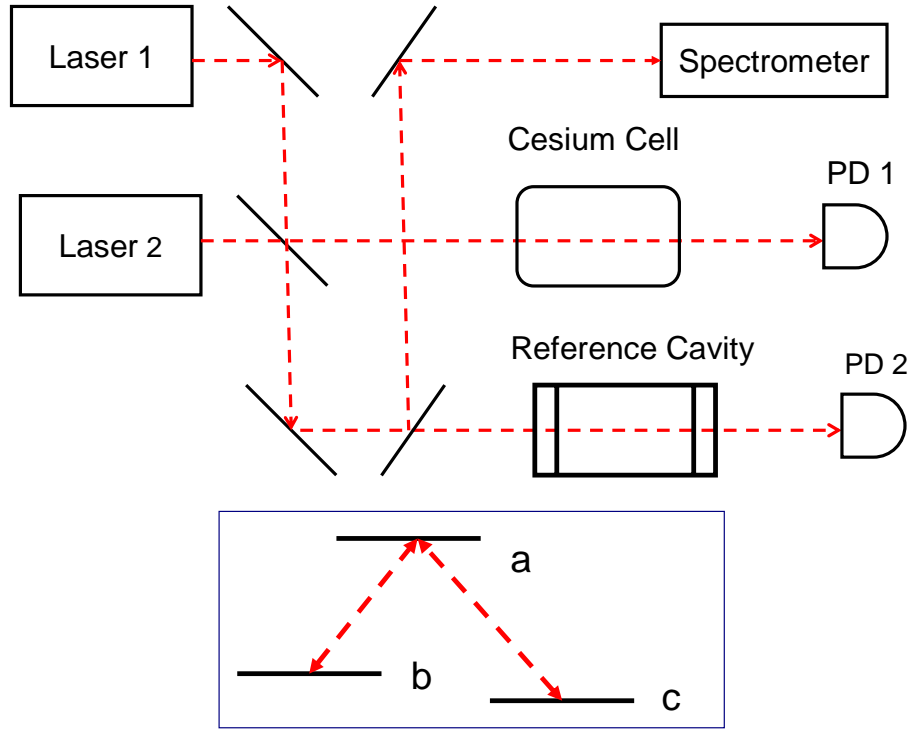

Figure 1. Schematic of the setup and a simplified energy level scheme.

Observed Λ-resonance at temperature 495 K and optical power of each beam 1.5 mW is shown in Fig.2. The width of EIT resonance is of 3.9 MHz. The frequency difference between two lasers is tuned around 62.2 GHz (wavelength difference is near 0.1 nm). By using our recorded absorption spectra and results of [12-15] we have got the rotational quantum numbers 89 and 90 of the ground states. We have determined the vibrational quantum numbers as 11 for the ground states and 6 for excited states.

4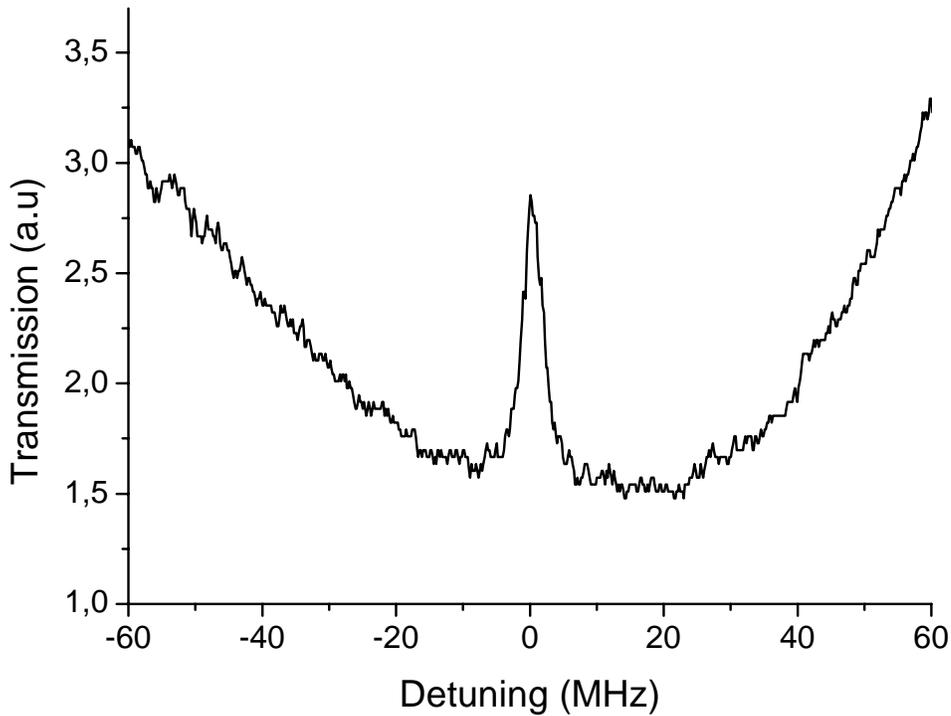

Figure 2. Transmission versus two-photon detuning.

The width of EIT resonance grows with the cell temperature. The dependence of EIT width on a total cesium vapor pressure is presented in fig. 3. The cesium vapor has atomic and molecular components which can be calculated by using the cell temperature T [16]. The pressure P and the number density N of ideal gas are related by well known expression $P = k*N*T$, where k is Boltzmann constant. The ratio of the molecular number density and atomic number density is order of $10^{-2}$ [16]. In the temperature range from 470K to 530K the molecular number density changes from $10^{13}$ to $10^{14}$ cm$^{-3}$, atomic number density changes from $10^{15}$ to $10^{16}$ cm$^{-3}$. In our case the broadening mechanism of $\Lambda$-resonances is not clear yet, therefore we have plotted the EIT width versus the total cesium vapor pressure. The measured EIT width is changed more than five times, from 1.9 MHz at 11.5 Pa to 10 MHz at 124 Pa. From the fit of experimental data by linear function $W = W_0 + B*P$ we have got the slope $B = 0.066$ (0.004) MHz/Pa and a low power limit $W_0 = 1.6$ (0.2) MHz. We have estimated the contribution of the laser effective linewidth to the EIT width as 0.7 MHz by measuring beat note between two lasers. By taking into account residual Doppler broadening due to divergence and angle between laser beams [17] the estimated the low pressure limit of EIT width is less than 0.5 MHz.

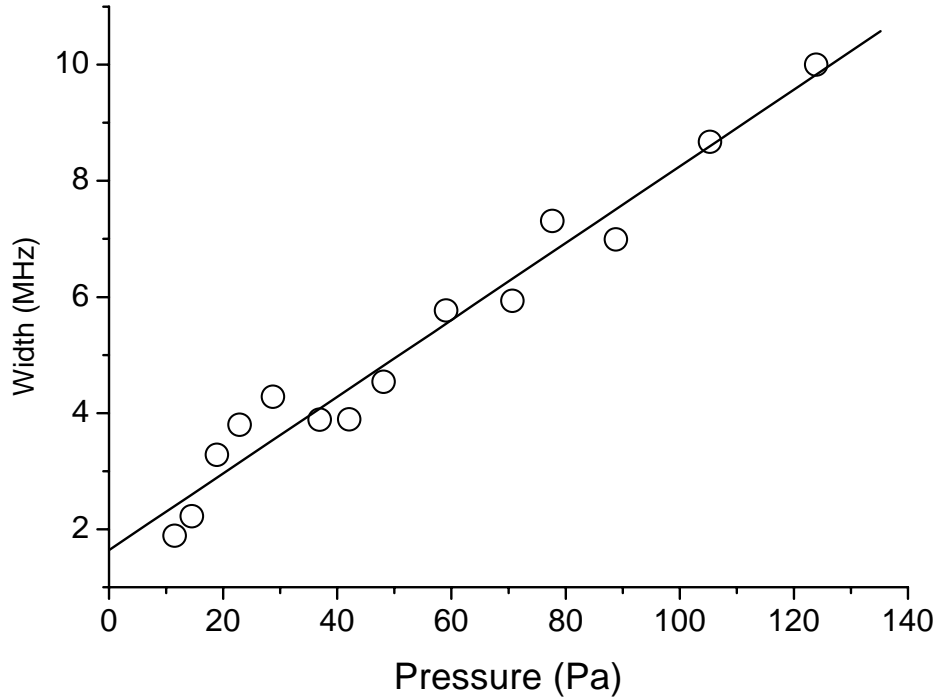

Figure 3. EIT width versus cesium vapor pressure.

Also we study the vapor pressure dependence of Doppler-free saturation resonances at single photon transitions (transitions $|b\rangle - |a\rangle$ or $|c\rangle - |a\rangle$, Fig.1). The widths of the saturation resonances are order of 20 MHz and vapor pressure induced variation of the spectral widths are less than 15%. Earlier similar pressure independence of the width of the saturation resonance in cesium molecules has been observed in [18]. The single-photon saturation of absorption has been associated with optical pumping (laser beam produced redistribution of population of the ground states) of the rovibrational ground states and the interaction time has been identified as diffusion time of the cesium molecules in surrounding cesium atomic gas inside the laser beam.

The different pressure dependence of the Doppler-free saturation resonance and narrow Λ-resonance could be attributed to different relaxation processes. In one photon transition only one rovibrational ground state is involved. Optical pumping process is dominated. In two-photon resonance two rovibrational ground states are involved and collision induced de-coherence between the states is responsible for broadening of the Λ-resonance. The de-coherence rate can depend on rotational and vibrational quantum numbers. It will be interesting to investigate Λ-resonances for various combinations of the rovibrational ground states. Also it will be useful to study EIT in super-sonic molecular beam where inter-particle interactions are strongly reduced [15]. Under these conditions decoherence time can be defined only by the interaction time and very narrow resonances can be observed.

The narrow Λ-resonances (Raman resonances) in molecules can be used as references for femtosecond laser frequency combs [19]. By using a train of short pulses the Λ-resonances were observed in rubidium atoms where hyperfine splitting of the ground state is just order of 0.01 nm [20, 21]. The spectral difference between allowed two-photon transitions between the rovibrational ground states in the cesium molecules can reach more than 40 nm in cesium molecules [12-15]. We assume that EIT in alkaline molecules produced by ultrashort pulses can be used to build up a compact time and frequency standard. It can be complimentary device to a HeNe-$CH_4$-based optical molecular clock [22]. In this work [22] a vibrational infrared transition (3.39 μm) has been used as a primary reference for the femtosecond optical frequency chain by using frequency conversion.

### 3. Conclusions

We have studied electromagnetically induced transparency (EIT) in cesium diatomic molecules in a vapor cell. We have observed narrow Λ-resonance in cesium molecular band $B^1\Pi_u$ - $X^1\Sigma_g^+$. The narrow EIT resonances can be used as frequency references for a femtosecond optical frequency comb.

**Acknowledgements**

We are thankful to P. R. Hemmer, D. R. Herschbach, R. L. Kolesov and G.R.Welch for useful discussions. This work is supported by the NSF (grant EEC-0540832) and  the Robert A. Welch Foundation (grant #A1261).

**References**

1. M. Fleishhauer, A. Imamoglu, J. P. Marangos: Rev. Mod. Phys. 77, 633 (2005).

2. L. Hau, S.E. Harris, Z. Dutton, C. H. Behroozi: Nature 397, 594 (1999).

3. M. M. Kash, V. A. Sautenkov, A. A. Zibrov, L. Hollberg, G. R. Welch, M. D. Lukin, Y. Rostovtsev, E. S. Fry, M. O. Scully: Phys. Rev. Lett. 82, 5229 (1999).

4. Y. V. Rostovtsev, Z.-E. Sariyanni, M. O. Scully: Phys. Rev. Lett. 97, 113001 (2006).

5. J. Kitching, S. Knappe, L. Hollberg: Appl. Phys. Lett. 81, 553 (2002).

6. P. D. D. Schwindt, S. Knappe, V. Shah, L. Hollberg, J. Kitching: Appl. Phys. Lett. 85, 6409 (2004).

7. J. B. Qi, A. M. Lyyra: Phys. Rev. A 73, 043810 (2006).

8. L. Li, P. Qi, A. Lazoudis, E. Ahmed, A. M. Lyyra: Chem. Phys. Lett. 403, 262 (2005).

9. S. Ghosh, J. E. Sharping, D. G. Ouzounov, A. L. Gaeta: Phys. Rev. Lett. 94, 093902 (2005).

10. R. Gupta, W. Happer, J. Wagner, E. Wennmyr: J. Chem. Phys. 68, 799 (1978).

11. M. Terrel, M. F. Masters: Am. J. Phys. 64, 1116 (1996).

12. M. Raab, G. Honing, W. Demtröder, C. V. Vidal: J. Chem. Phys. 76, 4370 - 4386 (1982).

13. J. Verges, C. Amiot: J. Mol. Spectroscopy, 126, 393 (1987).

14. U. Diemer, R. Duchowicz, M. Ertel, E. Mehdizadeh, W. Demtröder: Chem. Phys. Lett. 164, 419 (1989).

15. G. Brasen, W. Demtröder: J. Chem. Phys. 110, 11841 (1999).

16. A. N. Nesmeyanov: Vapor Pressure of Chemical Elements, R. Gary, ed. (Elsivier, 1963).

17. C. Bolkart, D. Rostohar, M. Weitz: Phys.Rev. A 71, 043816 (2005).

18. H. Chen, H. Li, Y. V. Rostovtsev, M. A. Gubin, V. A. Sautenkov, M. O. Scully: JOSA B, 23, 723 (2006).

19. T.W. Hänsch: Rev. Mod. Phys. 78, 129 (2006).

20. V. A. Sautenkov, Y.V. Rostovtsev, C. Y. Ye, G. R. Welch, O. Kocharovskaya, M. O. Scully: Phys. Rev. A 71, 063804 (2005).

21. L. Arissian, J.-C. Diels: Optics Commun. 264, 169 (2006).

22. S. M. Foreman, A. Marian, J. Ye, E.A. Petrukhin, M. A. Gubin, O.D. Mücke, F. N. C. Wong, E. P. Ippen, F. X. Kärtner: Opt. Lett. 30, 570 (2005).